\documentclass[aps,prd,preprint,superscriptaddress,nofootinbib]{revtex4-1}
\usepackage{chay,graphicx}

\newcommand{\fms}[1]{{#1}\!\!\!/}
\newcommand{\fmsl}[1]{{#1}\!\!\!\!/}
\newcommand{\n}{\overline{n}}

\newcommand{\mM}{\mathcal{M}}
\newcommand{\mO}{\mathcal{O}}
\newcommand{\mP}{\mathcal{P}}
\newcommand{\mR}{\mathcal{R}}
\newcommand{\mS}{\mathcal{S}}

\newcommand{\mL}{\mathcal{L}}

\newcommand{\mJ}{\mathcal{J}}
\newcommand{\nn}{\frac{\fms{\overline{n}}}{2}} 
\newcommand{\nnn}{\frac{\fms{n}}{2}}

\begin{document}

\title{Endpoint behavior of high-energy scattering cross sections}

\def\KU{Department of Physics, Korea University, Seoul 136-701, Korea} 
\def\CERN{Theory Division, Department of Physics, CERN, CH-1211 Geneva 23, Switzerland}

\author{Junegone Chay}
\email[E-mail:]{chay@korea.ac.kr}
\affiliation{\KU}
\author{Chul Kim}
\email[E-mail:]{chul.kim@cern.ch}
\affiliation{\CERN\vspace{0.5cm}} 

\begin{abstract} \baselineskip 3.0 ex 
In high-energy processes near the endpoint, there emerge new contributions associated 
with spectator interactions. Away from the endpoint region, these new contributions are suppressed compared to the leading contribution,
but the leading contribution becomes suppressed as we approach the endpoint and the new contributions become comparable. We present how
the new contributions scale  as we reach the endpoint and show that they are comparable to the 
suppressed leading contributions in deep inelastic scattering
by employing a power counting analysis. The hadronic tensor in deep inelastic
scattering is shown to factorize including the spectator interactions, and it can be expressed in terms of the lightcone distribution amplitudes
of initial hadrons. We also consider the contribution of the spectator contributions in Drell-Yan processes. Here the spectator interactions
are suppressed compared to double parton annihilation according to the power counting.
\end{abstract}

\maketitle

\baselineskip 3.0 ex 

\section{Introduction}

The standard results of high-energy scattering processes based on the operator product expansion  are consistent and work very well, but they
are expected to be modified near the endpoint where the Bjorken variable $x$ approaches 1. Since the available phase space
is restricted near the endpoint region, peculiar physical results arise and there has been a lot of theoretical interest 
in the endpoint behavior of high-energy scattering processes. 

The kinematic peculiarity near the endpoint $x\sim 1 -\Lambda/Q$ manifests two features which 
do not show up away from the endpoint region, where $Q$ is a large scale
and $\Lambda$ is the typical QCD scale  for hadron masses.
Firstly, the soft Wilson lines accompanied by collinear particles do not cancel completely, and the remnant is combined to 
produce soft functions. Extracting the soft part is 
crucial in factorization proof.  In deep inelastic scattering (DIS) near the endpoint, since the invariant mass of the final-state particles is 
$p_X^2 \sim Q^2 (1-x)$, 
spectator particles after hard scattering can be either soft or collinear to the final-state jets leaving no particles in the beam direction. 
In Drell-Yan (DY) process near the endpoint $\tau = Q^2/s \to 1$, there can be only 
soft final-state particles except a lepton pair due to the kinematic constraint. Comparing these two processes near the endpoint, 
the configurations of soft particles in the final states are different, causing different types of soft interactions, and the factorization 
proof near the endpoint is affected significantly by the soft parts.

Secondly, the contribution of spectator partons to the scattering cross section, which is subleading away from the endpoint
region, is not negligible near the endpoint and it should be included in the scattering cross section. It is not because the spectator contributions
are enhanced, but because the leading contribution is suppressed near the endpoint to become of the same order as the spectator contribution.
The proof that the spectator interaction becomes also important and the factorization property including the spectator interactions are 
the main theme of this paper.
 
The momentum of an energetic hadron in the lightlike $n$-direction can be decomposed into 
\begin{equation} 
p^{\mu} = \n\cdot p \frac{n^{\mu}}{2} + p_{\perp}^{\mu} + n\cdot p \frac{\n^{\mu}}{2} = \mO (Q) + \mO(\Lambda) 
+ \mO(\Lambda^2/Q),
\end{equation} 
where the lightcone vectors $n^{\mu}$ and $\n^{\mu}$ satisfy $n^2=\n^2=0$ and $n\cdot \overline{n} =2$.
The hadron is constrained to be on the mass shell $p^2 \sim \Lambda^2$, so are the partons constituting the hadron, such that a scattering 
process can be described in terms of the parton distribution functions (PDF)
as the probability distribution. However, these constraints give rise to special kinematic situation near the endpoint. Since the 
active parton undergoing hard scattering carries most of the energy inside the hadron, the $n$-component of the momentum for the 
spectator partons is of order $\Lambda$.  These spectator partons can have momenta satisfying the relative scaling to be $n$-collinear, 
but they cannot be on the mass shell. 
If the spectator partons become soft with all the momentum components of order $\Lambda$, they can be on the mass shell. 
But the total momentum of the hadron,  being the sum of a collinear and a soft momenta, becomes
of order $P^2 \sim Q\Lambda$, which is far off mass shell. Therefore in order to be consistent with the constraints of the on-shellness at 
the partonic and at the hadronic levels, and the kinematic constraint in the endpoint region, the initial spectator quarks are energetic,
$n$-collinear, and undergo a large momentum transfer inside the hadron of order $Q^2$ or $Q\Lambda$. 

As a result, near the endpoint region, the spectator particles which are 
initially $n$-collinear become either $\n$-collinear or soft after the large momentum transfer of order $Q^2$ or $Q\Lambda$. 
This momentum transfer is related not to the hard scattering, but to the spectator interaction in the initial hadron.
This necessitates the spectator interaction with a large momentum transfer in the scattering process near the endpoint in order to 
reflect the kinematic restrictions consistently. 

As we will see later, it is the main reason for the suppression of the conventional scattering cross section 
\footnote{\baselineskip 3.0 ex 
We mean the `conventional scattering cross section' by the scattering cross section 
neglecting spectator partons.} near the endpoint, 
which becomes comparable to the contribution of the spectator interactions.
In the standard region $1-x \sim \mO(1)$, the spectator contribution with a large momentum transfer is suppressed by 
$\Lambda^2/Q^2$ compared to the leading conventional contribution, thus can be safely neglected. 
All-order factorization analyses (in $\alpha_s$) were presented in Refs.~\cite{Sterman:1986aj,Catani:1989ne} near the endpoint 
region for Drell-Yan processes, and the subleading contributions suppressed by powers of $\Lambda/Q$ from the final-state interactions 
via the subleading final-state jet functions were analyzed in Refs.~\cite{Chay:2005rz,Becher:2006mr,Akhoury:1998gs}. However, 
the issue of the spectator contribution has not been addressed in the limit $x\to 1$ in previous literature. Careful power 
counting indicates that the leading contribution obtained away from the endpoint region experiences severe suppression such that it is 
comparable to the spectator contributions as $x$ goes to 1.

DIS in the endpoint region has been so far conventionally described by the following schematic factorization 
formula~\cite{Sterman:1986aj,Catani:1989ne}
\begin{equation} 
\label{F1end}
F_1 (Q^2, x) \sim H (Q^2,\mu) \cdot J(Q^2 (1-x),\mu) \otimes f_{i/H} (x,\mu),
\end{equation} 
where $F_1$ is the conventional structure function in the endpoint, $H$ is a hard function, and $f_{i/H}$ is a PDF. 
$J$ represents the final-state jet function 
integrating out the degrees of freedom of order $Q^2(1-x)$. And `$\otimes$' denotes the convolution
of the jet function with the PDF. In the framework of soft-collinear 
effective theory (SCET)~\cite{Bauer:2000ew,Bauer:2000yr,Bauer:2001yt}, this factorization formula has been revisited and confirmed  
without considering the spectator interactions~\cite{Manohar:2003vb,Chay:2005rz,Becher:2006mr}. 
If the spectator contribution should be included near the endpoint as discussed above, 
the conventional leading contribution of Eq.~(\ref{F1end}) is to be modified including this contribution, too. The PDF includes both 
the collinear part in the beam and the soft part.  The collinear part can be described by the lightcone distribution amplitudes (LCDA) 
for the initial hadron, and the soft part includes the final-state soft spectator quarks, which modifies the structure of the PDF.

This mechanism  also affects the longitudinal structure function $F_L$. The dominant spectator contribution to $F_L$ comes from the 
subleading corrections to the current operator responsible for spectator interactions, and remarkably it becomes comparable to $F_1$, 
since $F_1$ is suppressed near the endpoint. The Callan-Gross relation states that $F_L = -F_1+ F_2 Q^2/(4x^2)$ vanishes 
at leading order in $1/Q$, but it does not have to hold at subleading order we consider here.  If we consider the subleading jet function related
to the final-state particles alone without the spectator contribution, it is shown that the contribution to $F_L$ is suppressed by 
$\Lambda/Q$~\cite{Chay:2005rz,Becher:2006mr,Akhoury:1998gs} compared to $F_1$. This arises from the subleading jet function 
by integrating out the degrees of freedom of order $p^2 \sim Q\Lambda$ in the final state. However the new contributions which 
will be considered here turn out to be dominant, compared to the contribution to $F_L$ from subleading jet functions without 
the spectator interaction.  

In Drell-Yan processes, the spectator particles can be in the original direction of the initial hadron as in DIS away from the endpoint region. 
Near the endpoint, since the invariant mass of the final-state hadrons is of order $\Lambda^2$, there can be only soft particles.
This is in contrast to DIS, since the spectator particles are either $\overline{n}$-collinear (collinear to the 
final-state energetic collinear particle) or soft in DIS near the endpoint. In DIS, the case with final 
$\overline{n}$-collinear particles corresponds to the endpoint limit of the conventional approach, and can be compared to the new 
contribution with the spectator interaction. But there is no such analog of the case with $\overline{n}$-collinear particles in DY processes. 
However, the situation gets more drastic since we should also consider the double parton annihilation 
in DY process, for it is less suppressed than the spectator interaction as far as the power counting is concerned. 

There is one hadronic scalar function to describe DY processes. The soft part differs from that in DIS, hence needs 
some modification  or a different definition in the PDF. Away from the endpoint region, the soft part cancels, and 
the PDF consists of the matrix elements of collinear operators. It enables us to use universal PDFs independent of the scattering processes. 
That is, if we obtain or define the PDF in DIS, it can be used in DY processes. Near the endpoint, the soft part does not cancel, and it 
should be included in the definition of the PDF.  If the PDF defined in DIS is to be
employed in DY processes, there should be some modification which incorporates the difference of the soft parts in the two processes. 

In this paper, we consider the new contributions arising from spectator interactions in DIS 
and DY processes. The power counting is performed systematically and it is shown that the size of the new contributions 
is comparable to the standard contribution near the endpoint and the factorization property is considered. 
In Section~\ref{power} we perform the power counting analysis in DIS in the large $x$ limit and show how the spectators 
engage in the scattering process. In Sec.~\ref{facdis}, we show the factorization property for the new contributions in DIS. We employ 
two-step matching to prove the factorization explicitly.  In Sec.~\ref{dyp}, we present the power counting analysis for Drell-Yan processes,
including the double parton annihilation.  In Sec.~\ref{conc}, we give a conclusion.

\section{Power counting in DIS as $x\rightarrow 1$  \label{power} }

A systematic power counting can be applied to study the suppression of the scattering cross section near the endpoint.  Let us 
illustrate how the power counting is performed in DIS first.
The momentum of the final states is given by $p_X = q + P$, where $q$ is the momentum transfer from the leptonic system and 
$P$ is the momentum of the initial hadron. The invariant mass of the final states is given by 
\begin{equation} 
p_X^2 = \frac{(1-x)}{x} Q^2 + m_H^2,
\end{equation} 
where $x = -q^2/2P\cdot q = Q^2/2P\cdot q$ and $m_H$ is the initial hadron mass. We now choose the Breit frame in which 
$q^{\mu} = Q(\n^{\mu} - n^{\mu})/2$ and the initial hadron is described as an $n$-collinear particle.
The invariant mass of the final state varies as $x$ changes. Away from the endpoint, $p_X^2 \sim Q^2$, and 
this represents general hard scattering processes. As $x\to 1$, the invariant mass gets smaller, and the limit is classified into two regions.
The first is the resonance region where $1-x \sim \Lambda^2/Q^2$ with
$p_X^2 \sim \Lambda^2$, in which  only $\n$-collinear hadrons are allowed kinematically in the final state. 
 And the second is the endpoint region $1-x \sim \Lambda/Q$ with $p_X^2 \sim Q\Lambda$, in which there can be $\n$-collinear jets
 and soft hadrons.

In both regions there are no $n$-collinear final-state particles, while the initial-state partons are $n$-collinear particles. Therefore 
the spectator particles have to interact with large momentum transfer to become either soft or $\overline{n}$-collinear, and we have to 
include all the  interactions of the initial partons. In the resonance region all the spectators undergo hard interactions with the 
momentum transfer of order $Q^2$, and then the spectator particles, which are initially $n$-collinear, are converted into $\n$-collinear 
particles to make $p_X^2 \sim \Lambda^2$.  In the endpoint region we have two possibilities: First, an $n$-collinear 
spectator inside the initial hadron can be $\n$-collinear undergoing hard interactions as in the resonance region. But here the 
offshellness of the final state is allowed to be of order $Q\Lambda$, much larger than the resonant case. Secondly, a spectator loses most of its 
energy to the active parton and becomes a soft particle. This energy transfer between the active parton and the spectator is 
hard-collinear in the $n$ direction. Its offshellness is of order $Q\Lambda$, which is the typical offshellness 
of the final-state jet in the endpoint region.  

Now we can perform the power counting of the hadronic tensor for inclusive DIS, which is defined as 
\begin{equation}
W^{\mu \nu} = \sum_X \int \frac{d^4 z}{2\pi} e^{iq\cdot z} \langle H|J^{\mu\dagger} (z) 
|X\rangle \langle X|J^{\nu} (0)|H\rangle,
\label{wmunu}
\end{equation}
where $H$ is the initial hadron, $J^{\mu}$ is an electromagnetic current, and the summation includes the phase space  of the 
final-state particles. The power counting on the volume $d^4 z$ depends on how much phase space 
is available. For power counting on the remaining part $\sum_X \langle H | J^{\mu\dagger} | X \rangle \langle X | J^{\nu} | H \rangle$, 
we divide it into three parts; the initial state, the amplitude squared, and the final state contributions.
Here we focus on the amplitudes at tree level, but the result on the power counting by $\eta \sim \Lambda/Q$ can be easily 
extended to loop corrections because no loop contribution can enhance the amplitude by inverse powers of $\eta$.

In the standard region where $1-x\sim\mO(1)$, $d^4 z$ covers the full phase space, hence power-counted as $1/Q^4$.
The initial-state part is schematically written as $|\langle 0 | \Psi_n | H \rangle |^2$. Here $\Psi_n$ is an $n$-collinear quark and 
scales as $Q^{3/2} \eta$ with $\eta \sim \Lambda/Q$, and the collinear state $|H\rangle$ scales as $1/\Lambda \sim 1/(Q\eta)$.
Therefore the initial-state part yields the factor $Q$. The final state contains $\int d^4 p \delta(p^2)  \FMslash{p}$, where 
$\FMslash{p}$ comes from the spinor sum of the final state. Because the final state carries the 
hard momentum in the standard region, the power counting states that 
$\int d^4 p \delta(p^2)  \FMslash{p} \sim Q^4 \cdot (1/Q^2) \cdot Q = Q^3$. 
Also the amplitude squared is simply $\mO(1)$. In the standard region, spectator contributions do no change 
the power counting since they are of order 1 or give higher powers of $\eta$.  Therefore the overall power counting 
for the structure function yields $\mO(1)$. 

As explained above, the spectator contribution should be included near the endpoint region. It is also important how many particles there are 
in the leading Fock space of the initial hadron $H$. In the case of a pion, there are $q\bar{q}$ in the leading Fock space and $qqq$ 
for a proton. And the power counting on the structure functions for a pion and a proton is different. Because all the partons are 
involved in the scattering process, the time-ordered products of the electromagnetic current and the interaction 
Lagrangians including all the spectators should be taken into account in the hadronic tensor $W^{\mu\nu}$. For the power 
counting of the initial-state contributions, 
we consider  $|\langle 0 | \overline{\Psi}_n \Psi_n | \pi\rangle |^2 $ for an initial-state pion and 
$|\langle 0|\Psi_n \Psi_n \Psi_n|p\rangle|^2$ for a proton neglecting irrelevant Lorentz structure and color factors. 
From our power counting rule, these yield the factors $Q^2\Lambda^2$ and $Q^{3}\Lambda^4$ respectively, and 
the structure function for the proton is more suppressed than the structure function for the pion near the endpoint.

\begin{figure}[b]
\begin{center} 
\includegraphics[height=5.5cm]{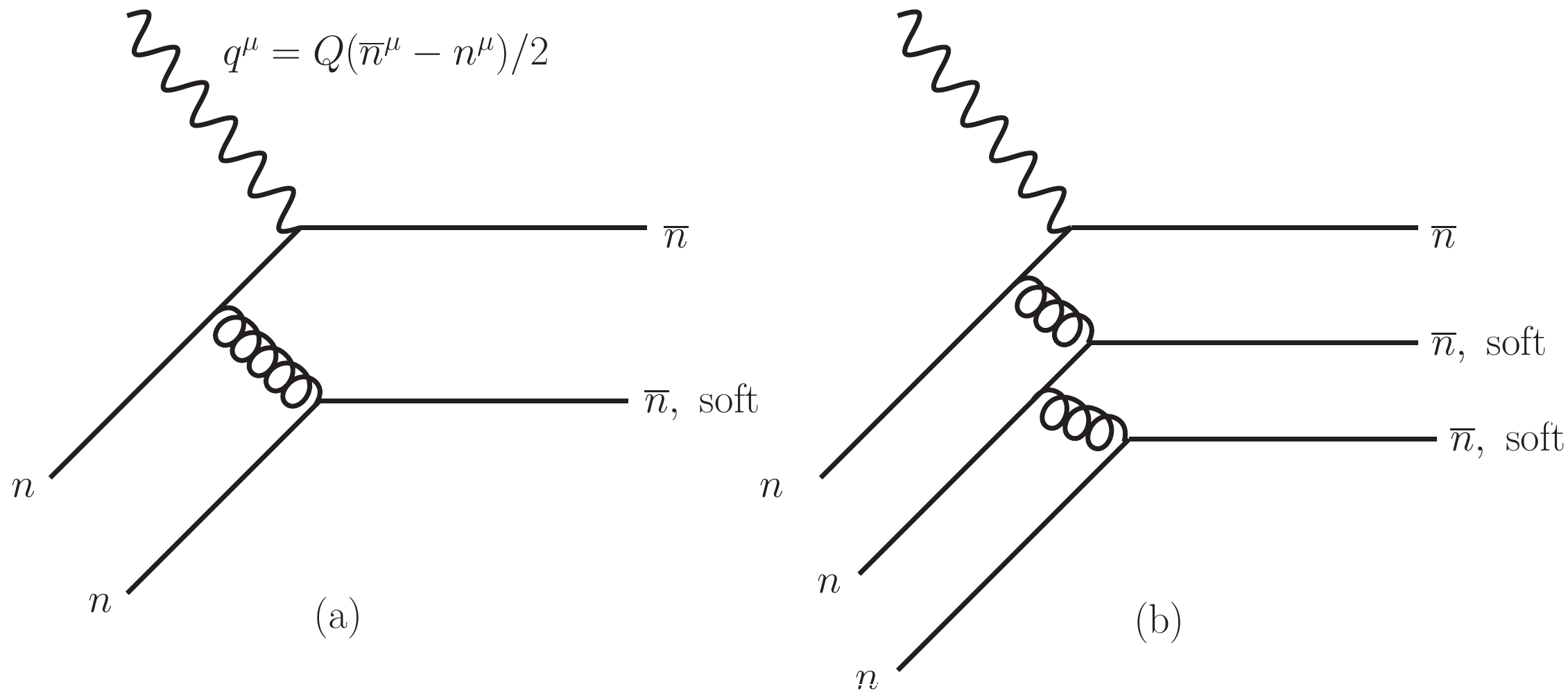}
\end{center}\vspace{-.5cm}
\caption{\baselineskip 3.0ex 
Specific examples of DIS processes (a) for a pion and (b) for a proton in the initial state near the endpoint. In the resonance region 
$1-x \sim \mO(\eta^2)$, all the final-state quarks are 
$\n$-collinear $(p_c^2 \sim \Lambda^2)$. In the endpoint region $1-x \sim \mO(\eta)$, the spectator quarks in the final state can 
be either $\n$-hard-collinear ($p_{hc}^2 \sim Q\Lambda$) or soft ($p_s^2 \sim \Lambda^2$).}
\label{spectatorcontrib}
\end{figure}

In the resonance region the phase space is severely constrained and the invariant mass of the final states becomes 
$p_X^2 \sim Q^2(1-x)+m_H^2 \sim \Lambda^2$.  The momentum $p_X$ flows between the two points 0 and $z$ 
in the hadronic tensor, and it implies that the volume $d^4 z$ is counted as $1/\Lambda^4$. 
Some examples of DIS near the endpoint for an initial pion and a proton are shown in  Fig.~\ref{spectatorcontrib}. 
The momentum transfer between the active and the spectator quark is hard ($p_h^2  \sim Q^2)$, hence the amplitudes for an initial  
pion and a proton scale as $1/Q^3$ and $1/Q^6$ respectively. Each quark field in the final state is power counted as 
$\int d^4 p_c \delta(p_c^2)  \FMslash{p}_c \sim \Lambda^4 \cdot (1/\Lambda^2) \cdot Q = Q\Lambda^2$, where $p_c$ 
represents collinear momentum with the offshellness of order $\Lambda^2$. Combining all the factors, the power counting 
of the hadronic tensor is given as 
\begin{eqnarray} 
W^{\mu\nu} &\sim& \mM^2 \cdot I \cdot F \cdot V, \nonumber \\
&\sim& \left\{ \begin{array}{ll}
\displaystyle \Bigl(\frac{1}{Q^3}\Bigr)^2 \cdot Q^2\Lambda^2 \cdot (Q\Lambda^2)^2 \cdot \frac{1}{\Lambda^4} \sim 
\frac{\Lambda^2}{Q^2} \sim 1-x  & \mbox{for} \  H = \pi, \\
\displaystyle \Bigl(\frac{1}{Q^6}\Bigr)^2 \cdot Q^3\Lambda^4 \cdot (Q\Lambda^2)^3 \cdot \frac{1}{\Lambda^4} \sim \frac{\Lambda^6}{Q^6}
\sim (1-x)^3  & \mbox{for}  \ H = p,
\end{array}
\right.
\end{eqnarray} 
where $\mM^2$ denotes the amplitude squared, $I(F)$ is the initial (final) state, and $V$ indicates 
the volume $d^4 z$. This is consistent with the previous power counting in the resonance region \cite{Blankenbecler:1974tm}.

In the endpoint region $p_X^2$ scales as $Q^2 (1-x) \sim Q\Lambda$ and $d^4z$ is counted as $1/(Q\Lambda)^2$. 
The spectator quarks in the final state can be either hard-collinear ($p_{hc}^2 \sim Q\Lambda$) or soft ($p_s^2 \sim \Lambda^2$), 
while the active parton in the final state is kept to be hard-collinear 
for the maximal scaling. We estimate the power counting of amplitudes from Fig.~\ref{spectatorcontrib}. For an initial pion the amplitude 
is power counted as either $1/Q^3$ (hard momentum transfer) or $1/(Q^2\Lambda)$ (hard-collinear momentum transfer). For an initial 
proton the amplitude is estimated to be of order $1/Q^6$ (two hard-collinear spectators), 
$1/(Q^5\Lambda)$ (one hard-collinear and one soft spectators), and $1/(Q^4\Lambda^2)$ (two soft spectators). The hard-collinear 
final state is maximally power counted as $\int d^4 p_{hc} \delta(p_{hc}^2)  \FMslash{p}_{hc} \sim Q^2\Lambda^2 
\cdot (1/Q\Lambda) \cdot Q = Q^2\Lambda$ and the final soft state scales 
as $\int d^4 p_{s} \delta(p_{s}^2)  \FMslash{p}_{s} \sim \Lambda^4 \cdot (1/\Lambda^2) \cdot \Lambda = \Lambda^3$. The final 
results of the scaling behavior of  $W^{\mu\nu}$ are summarized in Table.~\ref{table}. Near the endpoint, the hadronic tensor scales
as $(1-x)^2$ for an initial pion, and $(1-x)^5$ for an initial proton for all the possible final states. The point is that the suppression of the 
hadronic tensor near the endpoint is the same for the final $\overline{n}$-collinear and soft particles, and depends only on the type 
of the initial hadrons. 

\begin{table*}
\begin{tabular}{|c|c|c|c|c|c||c|}\hline
$H$  & Final spectators & $\mM^2$ & $I$ & $F$ & $V$ & $W^{\mu\nu}$ \\ \hline\hline
~~$\pi$~~ & $\Psi_{\n,hc}$ & ~$(1/Q^{3})^2$~& ~$Q^2\Lambda^2$~ & ~$(Q^2 \Lambda)^2$~ & ~$1/(Q^{2}\Lambda^{2})$~ 
& $~\sim (1-x)^2~$ \\ 
 & $q_s$ & $~(1/Q^{2}\Lambda)^2~ $ & ~$Q^2\Lambda^2$~ & $Q^2 \Lambda\cdot \Lambda^3 $ & $1/(Q^{2}\Lambda^{2})$ & 
$\sim (1-x)^2$ \\ \hline\hline
 & $\Psi_{\n,hc}$,$\Psi_{\n,hc}$ & ~$(1/Q^{6})^2$~& ~$Q^3\Lambda^4$~ & ~$(Q^2 \Lambda)^3$~ & ~$1/(Q^{2}\Lambda^{2})$~ 
& $~\sim (1-x)^5~$ \\ 
 ~~$p$~~& $\Psi_{\n,hc},~q_s$ & ~$(1/Q^{5}\Lambda)^2$~& ~$Q^3\Lambda^4$~ & ~$(Q^2 \Lambda)^2 \cdot \Lambda^3$~ 
& ~$1/(Q^{2}\Lambda^{2})$~ & $~\sim (1-x)^5~$ \\ 
& $q_s,~q_s$ & ~$(1/Q^{4}\Lambda^2)^2$~& ~$Q^3\Lambda^4$~ & ~$(Q^2 \Lambda) \cdot (\Lambda^3)^2$~ 
& ~$1/(Q^{2}\Lambda^{2})$~ & $~\sim (1-x)^5~$ \\ \hline
\end{tabular}
\caption{\baselineskip 3.0ex 
The scaling behavior of the hadronic tensor $W^{\mu\nu}$ in the endpoint region $(1-x\sim \Lambda/Q)$. $H$ is an initial hadron, 
$\Psi_{\n,hc}$ represents an $\n$-hard-collinear quark, and $q_s$ is a soft quark.}
\label{table}
\end{table*}

\section{Factorization analysis of DIS near the endpoint\label{facdis}}

In this section we analyze the factorization of DIS near the endpoint in SCET. 
For simplicity we consider DIS with an initial pion rather than with a proton. But the extension to the initial proton  is straightforward. 
The general tensor structure of  $W^{\mu\nu}$ for DIS in the Breit frame can be written as
\begin{equation}
W^{\mu\nu}=\Bigl( -g^{\mu\nu} +\frac{q^{\mu} q^{\nu}}{q^2} \Bigr) F_1  + \Bigl( P^{\mu}- \frac{P\cdot q}{q^2} q^{\mu} \Bigr) 
\Bigl( P^{\nu}-  \frac{P\cdot q}{q^2} q^{\nu} \Bigr) F_2  = -g_{\perp}^{\mu\nu} F_1 + v^{\mu} v^{\nu} F_L,  
\end{equation}
where $g_{\perp}^{\mu\nu} = g^{\mu\nu} - (n^{\mu} \overline{n}^{\nu} +\overline{n}^{\mu} n^{\nu})/2$,  and 
$v^{\mu} = (n^{\mu}+\n^{\mu})/2$. The longitudinal structure function $F_L$ is defined as
$F_L = -F_1 + F_2 Q^2/(4x^2)$. Away from the endpoint region 
$F_L$ is suppressed compared to $F_1$. But as $x$ goes to 1, $F_1$ is suppressed and $F_L$ becomes comparable to $F_1$. 
Both structure functions are influenced by the spectator interaction in the endpoint region $1-x \sim \Lambda/Q$.

\subsection{Contribution from  hard gluon exchange}

For the hard gluon exchange as shown in  Fig.~\ref{spectatorcontrib}, we obtain the local SCET operators with $n$ and $\n$-collinear 
quark fields by integrating out hard gluons. For an initial pion, these operators are obtained from  Fig.~\ref{spectatorcontrib} (a) 
along with the hard gluon exchange between the outgoing active quark and the spectator quark. After matching these contributions 
onto SCET, the electromagnetic current $J^{\mu}=\bar{q}\gamma^{\mu}q$ is expressed 
in terms of a convolution as
\begin{equation}
\label{convhard} 
v^{\mu} C_H  \otimes O_H = v^{\mu} \int^1_0 du dw~C_H (u,w,Q) O_H(u,w), 
\end{equation} 
where  $O_H (u,w)$ is given by
\begin{equation} 
\label{pihardO}
O_H (u,w) = \frac{1}{Q^3} \overline{\Psi}_{\n} \delta \Bigl(w-\frac{n\cdot \mR^{\dagger}}{Q} \Bigr) \tilde{Y}_{\n}^{\dagger} \gamma_{\perp}^{\alpha} T^a Y_n 
\delta \Bigl(u-\frac{\n\cdot \mP}{Q} \Bigr) \Psi_n \cdot 
\overline{\Psi}_{n} Y_n^{\dagger} \gamma^{\perp}_{\alpha} T^a \tilde{Y}_{\n} \Psi_{\n}.
\end{equation}
Here we take the active quark as a quark and the spectator quark as an antiquark. The SCET collinear field 
$\Psi_{n(\n)} = W_{n(\n)}^{\dagger} \xi_{n(\n)}$ is collinear-gauge invariant, where $W_{n(\n)}$ is an $n(\n)$-collinear Wilson 
line \cite{Bauer:2000yr}. The variables $u$ and $w$ are the momentum fractions of the active quark before and after the hard scattering
respectively, and $\n\cdot \mP$ ($n\cdot \mR$) is 
the label momentum of the $n(\n)$-collinear field. $C_H$ is the Wilson coefficient for the hard gluon exchange, and at tree level it is given by 
$C_H^{(0)} =8\pi\alpha_s/(\bar{u}\bar{w})$ with  $\bar{u}=1-u$ and $\bar{w}=1-w$. 

We have redefined the collinear quark fields to decouple soft interactions as $\Psi_{n(\n)} \to Y_{n(\n)} \Psi_ {n(\n)}$ (annihilated~quark) and 
$\Psi_{n(\n)} \to \tilde{Y}_{n(\n)} \Psi_ {n(\n)}$ (created~antiquark) in Eq.~(\ref{pihardO}), where the soft Wilson lines are defined as \cite{Bauer:2001yt,Chay:2004zn} 
\begin{eqnarray}
\label{softWn}
Y_{n} (x) &=& P\exp \Bigl[ig\int_{-\infty}^x ds n\cdot A_{\mathrm{s}} (ns)\Bigr],~~~\tilde{Y}_{n} (x) = 
\overline{P}\exp \Bigl[-ig\int_{\infty}^x ds n\cdot A_{\mathrm{s}} (ns)\Bigr] \\
\label{softWnbar}
Y_{\bar{n}} (x) &=& P\exp \Bigl[ig\int_{-\infty}^x ds \overline{n}\cdot 
A_{\mathrm{s}} (\overline{n}s)\Bigr],~~~\tilde{Y}_{\n} (x) = 
\overline{P}\exp \Bigl[-ig\int_{\infty}^x ds \n\cdot A_{\mathrm{s}} (\n s)\Bigr],
\end{eqnarray}
where $P$ and $\overline{P}$ represent path-ordering and antipath-ordering respectively. 
Though there is only an octet four-quark operator at tree level, there can be singlet operators at higher order in $\alpha_s$, and we
can take the appropriate color projection for a color-singlet pion. Note that the result in Eq.~(\ref{convhard}) is proportional to $v^{\mu}$. 
Thus the hard-gluon exchange for the pion contributes to the longitudinal structure function $F_L$.

If we take the matrix element of Eq.~(\ref{convhard}) between the initial pion and the final state $X$, the $n$-collinear part can be 
expressed in terms of the pion light-cone distribution amplitude (LCDA)~\cite{Lepage:1980fj} because there is no outgoing final $n$-collinear 
particle. Using the expression for the leading-twist LCDA in SCET~\cite{Bauer:2001cu,Bauer:2002nz}
\begin{equation}
\label{LCDA}
\langle 0 | \Bigl[\delta(u-\frac{\n\cdot \mP}{\n\cdot p_{\pi}}) \Psi_n\Bigr]^a_{\alpha} 
\Bigl[\overline{\Psi}_n\Bigr]^b_{\beta} | \pi (p_{\pi}) \rangle 
= \frac{i}{4} f_{\pi} \n\cdot p_{\pi} \frac{\delta^{ab}}{N} \Bigl(\nnn \gamma_5\Bigr)_{\alpha\beta} \phi_{\pi} (u), 
\end{equation}
we have 
\begin{equation}
\label{hardmat} 
\langle X | C_H \otimes O_H | \pi \rangle
= i\frac{f_{\pi}}{Q^2} \int^1_0 du dw H(u,w,Q^2) \phi_{\pi} (u) 
\langle X_{\n} | \overline{\Psi}_{\n} \delta(w-\frac{n\cdot \mR^{\dagger}}{Q}) \nnn \gamma_5 \Psi_{\n} |0 \rangle,
\end{equation} 
where $\phi_{\pi}$ is the leading-twist LCDA for the pion, $N$ is the number of colors, and 
$H$ is the hard factor given by $C_F C_H^{(0)}/2N$ at tree level. This expression can be generalized to include  higher-order $\alpha_s$ 
corrections. In Eq.~(\ref{hardmat}) we put $\n\cdot p_{\pi} = Q$ neglecting $\mO(1-x)$. The soft Wilson lines in Eq.~(\ref{pihardO}) 
cancel since the pion is a color singlet. By inserting Eq.~(\ref{LCDA}), the matrix element in Eq.~(\ref{hardmat}) can be explicitly given as
\begin{eqnarray}
\langle X|C_H \otimes O_H |\pi \rangle &=& \frac{1}{Q^2} \frac{if_{\pi}}{4N} \int_0^1 du dw H(u,w,Q^2) \phi_{\pi} (u) \\
&\times& \langle X_{\n}| \overline{\Psi}_{\n} \delta \Bigl(\omega -\frac{n\cdot \mathcal{R}^{\dagger}}{Q}\Bigr)
\gamma_{\perp}^{\alpha} \frac{\FMslash{n}}{2}\gamma_5 \gamma_{\perp \alpha} \tilde{Y}_{\n}^{\dagger} T_a Y_n Y_n^{\dagger} 
T_a \tilde{Y}_{\n} \Psi_{\n} |0\rangle, \nonumber 
\end{eqnarray}
from which the cancellation of the soft Wilson lines can be clearly seen.

From Eq.~(\ref{wmunu}), the contribution of the hard-gluon exchange to $F_L$ can be written as 
\begin{eqnarray}
\label{FLH1} 
&&F_L^H (Q^2,x) = (2\pi)^3\Bigl(\frac{f_{\pi}}{Q^2}\Bigr)^2 \int^1_0 du'dw' du dw H^{*}(u',w',Q^2) \phi_{\pi} (u') H (u,w,Q^2) 
\phi_{\pi} (u) \\
~~&\times& \sum_{X_{\n}} \delta(q+p_{\pi}-p_X) \langle 0 |
\overline{\Psi}_{\n} \delta(w'-\frac{n\cdot \mR}{Q}) \nnn \gamma_5 \Psi_{\n} | X_{\n} \rangle \langle X_{\n} |
\overline{\Psi}_{\n} \delta(w-\frac{n\cdot \mR^{\dagger}}{Q}) \nnn \gamma_5 \Psi_{\n} |0 \rangle,
\nonumber
\end{eqnarray} 
where  the final-state jet function $J^H_{\n}$ is defined as 
\begin{eqnarray} 
\label{jetH}
&&Q^2 \int \frac{d^4 p_X}{(2\pi)^4} \delta(q+p_{\pi}-p_X) J_{\n}^H (w,w',p_X^2) \\
&&=\sum_{X_{\n}} \delta(q+p_{\pi}-p_X) \langle 0 |
\overline{\Psi}_{\n} \delta(w'-\frac{n\cdot \mR}{Q}) \nnn \gamma_5 \Psi_{\n} | X_{\n} \rangle \langle X_{\n} |
\overline{\Psi}_{\n} \delta(w-\frac{n\cdot \mR^{\dagger}}{Q}) \nnn \gamma_5 \Psi_{\n} |0 \rangle.
\nonumber
\end{eqnarray}
The computation of $J_{\n}^H$ is straightforward. At lowest order in $\alpha_s$, the momentum fractions $w$ and $w'$ should be 
the same because there is no collinear gluon emission to change the final momentum fraction.  In this case $J_{\n}^H$ is given by 
\begin{equation} 
J_{\n}^{H,(0)} (w,w',p_X^2) = \delta(w-w') K_{\n}^{H,(0)}(w,p_X^2),
\end{equation}
with $K_{\n}^{H,(0)}(w,p_X^2) = \pi (1-w)$. 

Putting Eq.~(\ref{jetH}) into Eq.~(\ref{FLH1}), we finally obtain the factorization formula as
\begin{eqnarray}
\label{FLH2} 
F_L^H (Q^2,x) &=& \frac{1}{2\pi}\Bigl(\frac{f_{\pi}}{Q}\Bigr)^2 \int^1_0 du'dw' H^{*}(u',w',Q^2) \phi_{\pi} (u') 
\int^1_0 du dw H (u,w,Q^2) \phi_{\pi} (u)  \\
&\times& J^H_{\n} (w,w',Q^2(1-x)),\nonumber
\end{eqnarray} 
with $p_X^2 = Q^2(1-x)$. As seen from Eq.~(\ref{jetH}), $J_{\n}^H$ is the quantity of order 1, but it can include the logarithm of 
$\ln(Q^2(1-x)/\mu^2)$  at higher orders in $\alpha_s$. Therefore $F_L^H$ is power counted as 
$f_{\pi}^2/Q^2 \sim \Lambda^2/Q^2 \sim (1-x)^2$ because all the other quantities 
are of order 1. This power counting is consistent with the result in Table~\ref{table}.

\subsection{Contribution from hard-collinear gluon exchange} 

For hard-collinear gluon exchange,  we employ two-step matching procedure  
QCD$\to \mathrm{SCET}_{\mathrm{I}} \to \mathrm{SCET}_{\mathrm{II}}$ by integrating out 
hard ($p_h^2 \sim Q^2$) and hard-collinear ($p_{hc}^2 \sim Q\Lambda$) degrees of freedom in turn. In $\rm{SCET_I}$ we do not 
distinguish the hard-collinear and the collinear fields allowing the fluctuations of $Q\Lambda$, while we keep only the collinear 
and soft fields with the fluctuations of $\Lambda^2$ in $\rm{SCET_{II}}$ after integrating out the hard-collinear fields. 

At tree level the electromagnetic current operator $J^{\mu} = \overline{q} \gamma^{\mu} q$ in the 
full theory can be expanded in powers of $\lambda = \sqrt{\Lambda/Q}$ in $\mathrm{SCET}_{\mathrm{I}}$ as  
\begin{eqnarray} \label{scetj}
J^{\mu} &=& \overline{\Psi}_{\bar{n}} \tilde{Y}_{\bar{n}}^{\dagger} \gamma_{\perp}^{\mu} Y_n 
\Psi_n 
- \frac{\overline{n}^{\mu}}{Q} \overline{\Psi}_{\bar{n}} \tilde{Y}_{\bar{n}}^{\dagger}
 Y_n \FMSlash{\mathcal{P}}_{\perp} \Psi_n  
-\frac{n^{\mu}}{Q} \overline{\Psi}_{\bar{n}} \FMSlash{\mathcal{P}}_{\perp}^{\dagger} 
\tilde{Y}_{\bar{n}}^{\dagger}  Y_n\Psi_n  \nonumber \\
&-&\frac{2v^{\mu}}{Q} \overline{\Psi}_{\bar{n}} \tilde{Y}_{\bar{n}}^{\dagger} 
Y_n \FMSlash{B}_{n\perp} \Psi_n -\frac{2v^{\mu}}{Q} \overline{\Psi}_{\bar{n}} \FMSlash{B}_{\n\perp}\tilde{Y}_{\bar{n}}^{\dagger} 
Y_n \Psi_n 
+\mathcal{O}(\lambda^2), 
\end{eqnarray}
where $B_{n}^{\mu} =[W_n^{\dagger} iD_{n}^{\mu} W_n]$,  
$B_{\bar{n}}^{\mu} =[W_{\bar{n}}^{\dagger} iD_{\bar{n}}^{\mu} W_{\bar{n}}]$
are the gauge-invariant collinear gauge fields, and the derivative operators act only 
inside the bracket. The first term in Eq.~(\ref{scetj}) is the leading current operator, 
the remaining operators are of order $\lambda$. 

Now we consider the hard-collinear gluon exchange between the electromagnetic current and the spectator quark. The spectator 
interaction is described by the following soft-collinear Lagrangian~\cite{Beneke:2002ph,Pirjol:2002km}
\begin{eqnarray}
\label{Lsc1} 
\mathcal{L}_{sc}^{(1)} &=& \overline{\Psi}_n \fmsl{B}_n^{\perp} Y_n^{\dagger} q_{s} + \mathrm{h.c.}, \\
\label{Lsc2a}
\mathcal{L}_{sc}^{(2a)} &=& \overline{\Psi}_n \nn n\cdot B_n Y_n^{\dagger} q_{s} + \mathrm{h.c.}, \\
\label{Lsc2b}
\mathcal{L}_{sc}^{(2b)} &=& \overline{\Psi}_n \nn W_n^{\dagger} i\fmsl{D}_n^{\perp} W_n \frac{1}{\n\cdot \mP} 
\fmsl{B}_n^{\perp}  Y_n^{\dagger} q_{s} + \mathrm{h.c.},
\end{eqnarray}
where the superscripts $(i)$ in $\mL_{sc}$ denote the $\lambda^i$ suppression compared to the leading SCET Lagrangian. 

The contribution of the hard-collinear gluon exchange is described in terms of the 
time-ordered products of the electromagnetic current and the 
soft-collinear Lagrangians in $\rm{SCET_I}$. However when we go down to $\rm{SCET_{II}}$ 
after integrating out the hard-collinear degrees of freedom,  the power counting changes accordingly. 
The collinear momentum in $\rm{SCET_{II}}$ scales as 
$p^{\mu} = (\n\cdot p, p_{\perp}, n\cdot p) = Q(1,\eta,\eta^2)$ with $\eta =\lambda^2$. 
The power counting of the $n$-collinear fields and their derivatives, $\mP_{\perp}$ and 
$n\cdot \mP$ changes $(\Psi_n,\mP_{\perp},n\cdot\mP) \sim (\lambda,\lambda,\lambda^2) 
\to (\eta,\eta,\eta^2)$ when matched onto $\rm{SCET}_{II}$. 

This fact implies that the final power-counting in $\rm{SCET_{II}}$ can be different from the power counting in $\rm{SCET_I}$.
An example is a hard-collinear gluon exchange in $\rm{SCET_I}$ from the time-ordered product of the leading electromagnetic 
current in Eq.~(\ref{scetj}) and $\mL_{sc}^{(1)}$, with the leading collinear Lagrangian $\mathcal{L}_c^{(0)}$. $\mathcal{L}_c^{(0)}$ 
contains an operator with two $D_n^{\perp}$'s, from which one hard-collinear gluon is contracted with $\mathcal{L}_{sc}^{(1)}$, 
and $\mathcal{P}_{\perp}$ is selected from another.  The resultant operator in $\rm{SCET_{II}}$ has a $\mP_{\perp}$, acting on 
the external $n$-collinear field\footnote{\baselineskip 3.0ex The derivative operator, 
$\mP_{\perp}$ does not vanish unless it returns a total transverse momentum of the pion. The nonvanishing $\mP_{\perp}$ contributes 
to twist-3 LCDAs if we take the matrix 
elements for the pion~\cite{Kim:2008ir}.} and it is suppressed by $\lambda$. Therefore this 
contribution is eventually power-counted as the same order as the operators from subleading time-ordered products, which  include neither 
$\mP_{\perp}$ nor $n\cdot\mP$. As a result, we include all the subleading contributions in the time-ordered products in 
$\rm{SCET_I}$ and the spectator contributions are given as
\begin{eqnarray} 
\label{timeo}
T_1^{\mu} &=& i\int d^4 x ~T\{J^{(0)\mu}_{\perp}(0), \mL_{sc}^{(1)} (x)\},~~~T_2^{\mu} 
= i\int d^4 x ~T\{J^{(0)\mu}_{\perp}(0), \mL_{sc}^{(2a)} (x)\}, \\ 
T_3^{\mu} &=& -\int d^4 x d^4 y~ T\{J^{(0)\mu}_{\perp}(0), \mL_{sc}^{(1)} (x), \mL_{c}^{(1)} (y)\},~~~
T_4^{\mu} = i \int d^4 x~ T\{J_L^{(1)\mu} (0), \mL_{sc}^{(1)} (x)\}, \nonumber 
\end{eqnarray} 
where $J_{\perp}^{(0)\mu}=\overline{\Psi}_{\bar{n}} \tilde{Y}_{\bar{n}}^{\dagger} \gamma_{\perp}^{\mu} Y_n 
\Psi_n$ and $J_L^{(1)\mu}=-(2v^{\mu}/Q) \overline{\Psi}_{\bar{n}} \tilde{Y}_{\bar{n}}^{\dagger} 
Y_n \FMSlash{B}_{n\perp} \Psi_n$, which are the first and the fourth operators in Eq.~(\ref{scetj}). $\mL_c^{(1)}$ is the 
subleading collinear Lagrangian given by~\cite{Chay:2002vy} 
\begin{equation}
\mathcal{L}_c^{(1)} = \overline{\Psi}_n Y_n^{\dagger} i\FMSlash{D}^{\perp}_s Y_n \frac{1}{\overline{n}\cdot \mathcal{P}} 
W_n^{\dagger} i\FMSlash{D}_{n}^{\perp} W_n \Psi_n + \mathrm{h.c.}.
\end{equation}

The fifth operator in Eq.~(\ref{scetj}) describes the interaction of $\overline{n}$-collinear particles, that is, the jet function, and it
has been considered to give a dominant contribution to the longitudinal structure function $F_L$, while its overall contribution is 
suppressed compared to $F_1$~\cite{Chay:2005rz,Becher:2006mr,Akhoury:1998gs}. But it turns out that $J_L^{(1)\mu}$ is another 
source for $F_L$, and this contribution to $F_L$ is comparable to the suppressed $F_1$ near the endpoint region in the power counting  
from the above analysis.

Our approach to the leading contribution in Eq.~(\ref{timeo}) is similar to the analysis for the heavy-to-light current for $B \to \pi$ or 
$K$ transition in semileptonic $B$ decays~\cite{Bauer:2002aj,Manohar:2006nz}, where the leading and subleading current operators 
involving a collinear gluon give comparable contributions in the power counting of $1/m_b$. The leading current obeys 
the heavy-to-light spin symmetry~\cite{Beneke:2000wa}, but the matrix element for the time-ordered 
products is nonfactorizable. It also has an endpoint divergence~\cite{Bauer:2002aj} or large ambiguities~\cite{Manohar:2006nz}. The remedy
for this problem is to absorb the nonfactorizable contributions to the form factor. 
However the contribution of subleading currents violates the spin symmetry, but it is factorizable. 
In DIS with an initial pion, 
$J_{\perp}^{(0)}$ and $J_L^{(1)}$ have 
also different spin structures. In a similar manner  the time-ordered products of $J_{\perp}^{(0)}$ have endpoint divergences if we 
take LCDAs for the pion, and they are absorbed into the nonperturbative hadronic matrix element,
while the time-ordered products of $J_{L}^{(1)}$ give factorizable contributions, and are free of endpoint divergence.

Including the radiative corrections, the relevant electromagnetic current operators can be written as
\begin{eqnarray} \label{j01}
J_{\perp}^{(0)\mu} &=& C_1(Q,\mu) \overline{\Psi}_{\bar{n}} \tilde{Y}_{\bar{n}}^{\dagger} 
\gamma_{\perp}^{\mu} Y_n \Psi_n,  \nonumber \\
J_L^{(1)\mu} &=& -\frac{2v^{\mu}}{Q} \int du~C_L (Q, u,\mu)  \overline{\Psi}_{\bar{n}} 
\tilde{Y}_{\bar{n}}Y_n \FMSlash{B}_{n\perp} \delta \Bigl( u- \frac{\overline{n}\cdot p}{Q}\Bigr)
\Psi_n,
\end{eqnarray}
where $C_1(Q,\mu)$ and $C_L (Q,u,\mu)$ are the Wilson coefficients. The Wilson coefficient $C_1(Q,\mu)$ has been computed 
to one loop \cite{Manohar:2003vb}.   Note that $C_L (Q, u,\mu)$ depends on the momentum fraction $u$ 
of the incoming quark because $J_L^{(1)}$ is a three-particle operator.
The anomalous dimension of $C_L$ is given by Eq.~(C8) in Ref.~\cite{Chay:2005rz}
to one loop. The renormalization group behavior of $J_{\perp}^{(0)}$ and $J_L^{(1)}$ is different since they are not a reparameterization-invariant 
combination \cite{Marcantonini:2008qn}. 

Schematically both the structure functions $F_1$ and $F_L$ can be written as
\begin{eqnarray} 
\label{F1s}
F_1  &\sim& H_1 \cdot \mathcal{J}_{\n} \otimes  K_1, \\
\label{Fls}
F_L^{HC}  &\sim& H_L \otimes \mathcal{J}_{\n} \otimes \mathcal{J}_L \otimes \mS_L \otimes \Phi_L,
\end{eqnarray}
where $H_1=|C_1(Q)|^2$, $H_L=C_L(Q,u) C_L^* (Q,v)$ are the hard factors, $\otimes$ denotes an appropriate convolution. $K_1$ is the
hadronic matrix element of collinear and soft operators, which come from $T_1^{\mu}$, $T_2^{\mu}$ and $T_3^{\mu}$. 
The contributions from $T_1^{\mu}$, $T_2^{\mu}$  and $T_3^{\mu}$ can be written in the form 
$\mathcal{J}_1 \otimes \mathcal{S}_1 \otimes \Phi_1$, but they contain endpoint divergences. On the other hand, 
$F_L^{HC}$ from $T_4^{\mu}$ is factorized. We put the superscript `$HC$' on $F_L$ to distinguish it from $F_L^H$, the hard 
gluon contribution to $F_L$. There are two kinds of jet functions $\mathcal{J}_{\n} (x,\mu)$, and $\mathcal{J}_{1,L} (x,\mu)$, obtained  
by integrating out the degrees of freedom of order $p^2 \sim Q\Lambda$ in the $\overline{n}$ and $n$ directions respectively. 
Physically $\mathcal{J}_{\n} (x,\mu)$ describes the final states, while $\mathcal{J}_{1,L} (u,\mu)$ describes the initial states of 
collinear particles. $\mS_{1,L}$ are the soft functions including soft Wilson lines and soft spectator quarks, and $\Phi_{1,L}$ are the 
LCDAs squared of the pion. 

In the conventional approach without including the spectator quark, 
the structure function $F_1$ can be cast into the following factorized form 
\begin{equation} \label{F1c}
F_1 (Q^2,x) =  H_1 (Q^2,\mu) \int dl \mathcal{J}_{\n} (Q(1-x)-l,\mu) f_{q/\pi}\Bigl(\frac{\n\cdot p_H-l}{\n\cdot p_H}\Bigr),
\end{equation}
where $f_{q/\pi}$ is the standard PDF obtained from the matrix element of a gauge-invariant collinear quark bilinear operator, 
\begin{equation}
\label{qpdf}
f_{q/H} (y) = \langle H | \overline{\Psi}_n \nn \delta (y \n\cdot p_{H} - \n\cdot\mP) \Psi_n |H \rangle,
\end{equation}  
and $H_1$ and $\mJ_{\n}$ are given in Eq.~(\ref{F1s}). The PDF can be additionally factorized into the soft and $n$-collinear parts, 
the combination of which recovers the renormalization behavior of the PDF~\cite{Chay:2005rz,Idilbi:2007ff}. 

The factorization formula, Eq.~(\ref{F1c}), holds even when the spectator contributions are included. It can be achieved 
if we generalize the definition of $f_{q/\pi}$ with the spectator contribution $K_1$. 
That is justified because the spin structure is the same for both contributions, and furthermore the renormalization behavior is also the same.
Note that the structure function $F_1$ is scale independent and the remaining parts in both 
the expressions of Eqs.~(\ref{F1s}) and (\ref{F1c}) are the same, therefore the renormalization group behaviors of $f_{q/\pi}$ and $K_1$ 
are also the same. In other words the spectator quark contributions involved in $K_1$ are described by $\mL_{sc}$ in $\rm{SCET_I}$, 
which is scale independent and does not affect the renormalization behavior. Therefore we can safely generalize $f_{q/\pi}$  to $K_1$   
without inducing additional complications, and $K_1$ or the PDF can be treated as a nonperturbative function to be determined from 
experimental data. As a result the definition of the standard PDF is still applicable near the endpoint region, but it holds up 
to $\rm{SCET_I}$. If we go further and employ the two-step matching, when the PDF is matched onto $\rm{SCET_{II}}$ including 
the spectator contributions,  it has more complicated substructure involving the lightcone distribution amplitudes of the initial hadrons. Note that 
$K_1$ or the PDF can be dependent on the scattering processes especially due to the difference of the soft functions in each scattering process. 
Theoretically the two-step matching result is more explicit, but it is not economical to express a 
nonperturbative quantity $K_1$ in terms of the convolutions of other nonperturbative quantities such as the LCDAs. 

For $F_L^{HC}$, we introduce a new nonperturbative function $f_{L}$ to cover $\mJ_L \otimes \mS_L \otimes \Phi_L$ in 
Eq.~(\ref{Fls}). Note that  the renormalization behavior of $f_L$ is different from $f_{q/\pi}$ because $H_1$ and $H_L$ have different 
anomalous dimensions. So $f_L$ is not related to $f_{q/\pi}$, and it is a new contribution to the PDF near the endpoint region. As we notice 
in the case of heavy-to-light transition 
in $B$ decays, $f_L$ can be factorized without the endpoint divergence. 

\begin{figure}[t]
\begin{center}
\includegraphics[height=4.3cm]{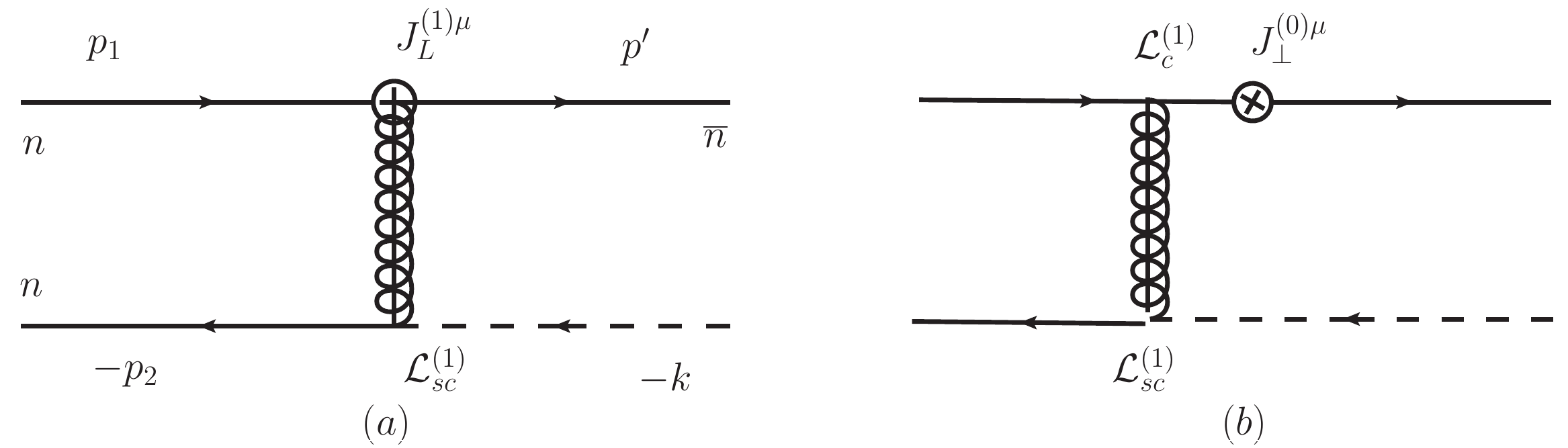} \vspace{-.5cm}
\end{center}
\caption{\baselineskip 3.0ex
Examples of the time-ordered products for the hard-collinear gluon exchange. Diagram (a) denotes 
$T_4^{\mu}$ and (b) describes $T_3^{\mu}$ in Eq.~(\ref{timeo}). 
The solid lines are collinear fermions, the dotted line  denotes an usoft quark, and the wiggly line with a solid line is an $n$-hard-collinear 
gluon with $p^2 \sim Q\Lambda$. \vspace{-.3cm}}
\label{topdis1}
\end{figure}

Now we consider the factorization proof for $F_L$ in detail in order to see how the spectator contributions can be treated in the 
inclusive scattering process. The first step is to compute the hard-collinear gluon exchange and construct a four-quark operator consisting 
of two incoming collinear quarks, an outgoing collinear quark and a soft quark. This four-quark operator with a soft quark can be obtained 
by the time-ordered product, $T_4^{\mu}$ in Eq.~(\ref{timeo}). The corresponding Feynman diagram is shown in Fig.~\ref{topdis1} (a). 
After integrating out the hard-collinear gluon at tree level, $T_4^{\mu}$ is written as
\begin{eqnarray} \label{t1}
T_4^{\mu} &=& 8\pi  \alpha_s  \frac{v^{\mu}}{Q^2} \int du \frac{C_L (u)}{\overline{u}} \int \frac{d\eta}{\eta} J_L (\eta)  \\
&\times& \overline{\Psi}_{\bar{n}} \tilde{Y}_{\bar{n}}^{\dagger} Y_n T^a \gamma_{\perp}^{\alpha} \Psi_n \cdot 
\overline{\Psi}_n \gamma^{\perp}_{\alpha} T^a \delta (\eta+n\cdot i\partial)
Y_n^{\dagger} q_{\mathrm{s}}, \nonumber 
\end{eqnarray}
where $\overline{u} =1-u$. And $J_L$ is the jet function obtained by integrating out the hard-collinear gluon in the $n$ direction, with 
the normalization $J_L (\eta)= 1 +\mathcal{O}(\alpha_s)$. 
At higher orders in $\alpha_s$, there can be a color singlet four-quark operator with the structure ${\bf 1 \otimes 1}$. Since 
the initial pion is a color singlet, we take the appropriate color projection. 
The matrix element of $T_4^{\mu}$ is given by
\begin{eqnarray}
\langle X|T_4^{\mu} |\pi\rangle &=& i  v^{\mu} \frac{4\pi C_F\alpha_s}{N}  \frac{f_{\pi}}{Q}  
\int \frac{du}{\overline{u}} C_L (u) \phi_{\pi} (u)   \\
&\times& \int\frac{d\eta}{\eta}  J_L (\eta) 
\langle X|\overline{\Psi}_{\bar{n}} \frac{\FMslash{n}}{2} \gamma_5 \tilde{Y}_{\bar{n}}^{\dagger} Y_n \delta (\eta+in\cdot \partial) 
Y_n^{\dagger} q_{\mathrm{us}} |0\rangle, 
\nonumber
\end{eqnarray}
where $\phi_{\pi}$ is the leading twist pion LCDA in Eq.~(\ref{LCDA}).

\begin{figure}[t] 
\begin{center}
\includegraphics[height=4.5cm]{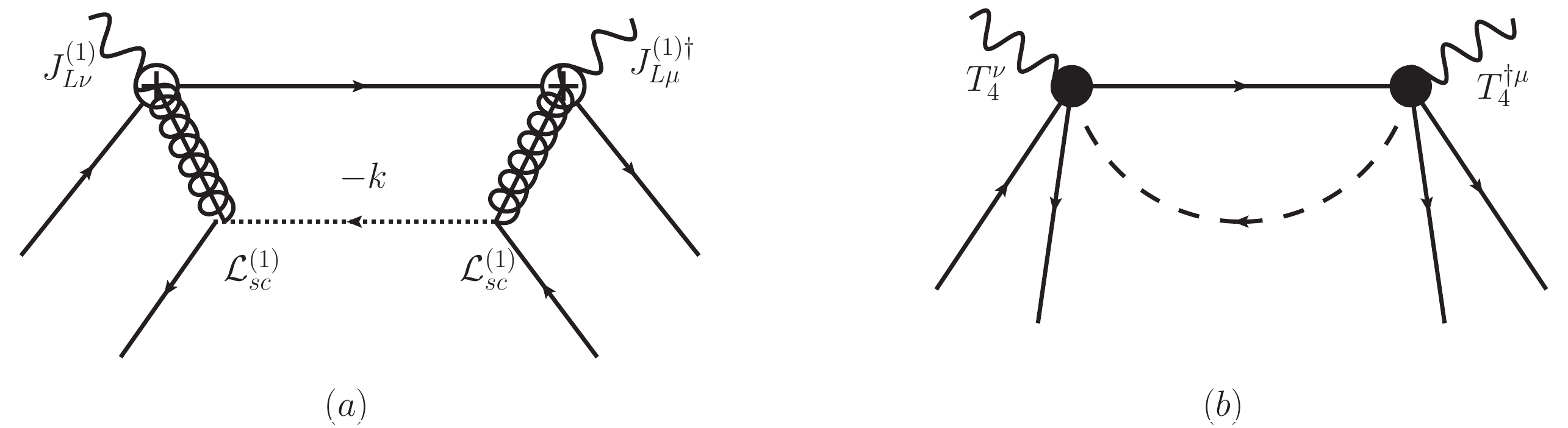}
\caption{\baselineskip 3.0ex
The Feynman diagrams for the longitudinal structure function $F_L$ with the hard-collinear gluon exchange. (a) The exchanged 
hard-collinear gluon in the $n$ direction is shown. 
(b) The equivalent diagram to (a) in terms of the time-ordered product of $T_4^{\dagger\mu}$ and $T_4^{\nu}$.  
\vspace{-.5cm}}
\label{subdis}
\end{center}
\end{figure}

The contribution of the hard-collinear gluon exchange to $F_L$ is obtained by replacing $J^{\mu}$ by $T_4^{\mu}$ in Eq.~(\ref{wmunu}), 
and the corresponding Feynman diagrams with $J_L^{(1)\mu}$ before integrating out the hard-collinear gluon and with $T_4^{\mu}$ 
are shown in Fig.~\ref{subdis} (a) and (b) respectively. The discontinuity of the Feynman diagrams in Fig.~\ref{subdis} yields the structure 
function. As a result the factorized contribution to $F_L$ is written as
\begin{equation}
\label{flhc}
F_L^{HC} (Q^2,x,\mu) = 2\frac{f_{\pi}^2}{Q^2} \int du dv T_L(u,v,\mu) \phi_{\pi} (u,\mu) \phi_{\pi} (v,\mu), 
\end{equation}
where the kernel $T_L(u,v,\mu)$ is given by
\begin{equation}
\label{tl}
T_L(u,v) = \frac{8\pi \alpha_s^2 C_F^2}{N^2} \frac{H_L (Q,u,v)}{\overline{u}\overline{v}}
\int dl \mJ_{\n} \Bigl(Q(1-x)-l\Bigr)
\int \frac{d\eta}{\eta} \frac{d\eta'}{\eta'} \mJ_L (\eta,\eta') \mS_L(l,\eta,\eta').
\end{equation}
In the case where the active quark is an antiquark and the spectator is a quark, the contribution is the same because of the charge symmetry. 
So we put the factor 2 in Eq.~(\ref{flhc}) reflecting this fact.  
Here the initial jet function is given by $\mJ_L (\eta,\eta') = J_L (\eta) J_L^*(\eta')$.
The soft function $\mS_L (l,\eta,\eta')$, which consists of soft quarks and soft Wilson lines,  is written as
\begin{equation} \label{soft}
\mS_L(l,\eta,\eta') = \langle 0| \overline{T}\Bigl[ \overline{q}_{s} Y_n \delta (\eta' -n\cdot i\overleftarrow{\partial}) 
Y_n^{\dagger} \tilde{Y}_{\bar{n}} \Bigr]\frac{\FMslash{n}}{2}  
\delta (l+\overline{n}\cdot i\partial)  T\Bigl[ \tilde{Y}_{\bar{n}}^{\dagger} Y_n \delta (\eta+n\cdot i\partial) Y_n^{\dagger} q_{s} 
\Bigr]|0\rangle,
\end{equation}
where $\overline{T}$ denotes the anti-time ordering.
The discontinuity of the soft quark propagator in the soft function in Eq.~(\ref{soft}) gives the factor
\begin{equation}
\int \frac{d^4 k}{(2\pi)^4} 2\pi \delta (k^2) \FMslash{k},
\end{equation}
from which the soft function at leading order in $\alpha_s$ is written as 
\begin{equation}
\mS_L^{(0)} (l,\eta,\eta') = \int \frac{d^4 k}{(2\pi)^3} \delta (k^2) 2n\cdot k \delta (\eta' -n\cdot k) \delta (\eta -n\cdot k) 
\delta (l-\overline{n}\cdot k)  
= \frac{1}{16\pi^3} l\delta (\eta-\eta').
\end{equation}
Note that the soft function is defined to be dimensionless. Definitely this is different from the soft function appearing in
the conventional approach which consists of only soft Wilson lines. The presence of  soft quarks gives a different soft function.
And the final jet function $\mJ_{\n} (\overline{n}\cdot p_{X_{\n}})$ with 
$\overline{n}\cdot p_{X_{\n}} = Q(1-x) -\overline{n}\cdot p_{Xs}$ ($p_{Xs}$ being the total momentum of the soft particles) is obtained 
from the relation
\begin{equation}
\label{fjet}
\sum_{X_{\bar{n}}} \langle 0 | \Psi_{\bar{n}}|X_{\bar{n}}\rangle \langle X_{\bar{n}}| \overline{\Psi}_{\bar{n}} |0\rangle = \int \frac{d^4 p_{X_{\bar{n}}}}{(2\pi)^4} 
\frac{\FMslash{\overline{n}}}{2} J(\overline{n} \cdot p_{X_{\bar{n}}}). 
\end{equation}
In this notation, the jet function at tree-level is given by $J(\eta) =2\pi \delta (\eta)$ and it has been computed to two-loop 
order \cite{Becher:2006qw}. 

Since $f_{\pi} $ is $\mO(\Lambda)$, $F_L^{HC}$ in Eq.~(\ref{flhc}) is power-counted as $\eta^2 \sim (1-x)^2$ as we expected in 
Table~\ref{table}. From Eqs.~(\ref{flhc}) and (\ref{tl}), the new nonperturbative function $f_L$ reads 
\begin{equation}
f_L (u,v, \mu) = \frac{8\pi f_{\pi}^2}{Q^2} \frac{\alpha_s^2 C_F^2}{N^2} \frac{\phi_{\pi} (u,\mu)}{\overline{u}} 
\frac{\phi_{\pi}(v,\mu)}{\overline{v}} \int \frac{d\eta}{\eta} 
\frac{d\eta'}{\eta'} \mJ_L (\eta,\eta',\mu)  \mS_L(l,\eta,\eta',\mu).
\end{equation}
Because $H_L$ and $\int dl \mJ_{\n}$ are of order 1, $f_L$ is also power-counted as order $\eta^2$. The same reasoning leads to 
the fact $f_{q/\pi} \sim \eta^2 \sim (1-x)^2$ because $W^{\mu\nu} \sim F_1 \sim (1-x)^2$ . 

The result can be extended to the case with an initial proton in a straightforward way, but it is definitely more complicated because 
there are more spectator quarks.  If we consider the similar factorization formulae for the structure functions  $F_{1,L} 
\sim  H_{1,L}~ (\times~\mathrm{or}~\otimes)~\mJ_{\n} \otimes f_{1,L}$, we can do the power counting on the 
nonperturbative functions $f_{1,L}$. Because $W^{\mu\nu} \sim F_{1,L} \sim \eta^5 \sim (1-x)^5$ as seen in Table~\ref{table} 
and $\mJ_{\n}$ is identical with the one defined in Eq.~(\ref{fjet}), both the nonperturbative functions $f_1$ and $f_L$ scale as $(1-x)^5$. 
From Ref.~\cite{Alekhin:2005gq} we can read off the fitted scaling behavior of the PDF from DIS experiments. At the factorization scale 
$\mu_F = 3~\rm{GeV}$ the powers of $(1-x)$ in the PDFs read $\sim 4$ for the $u$ valence quark and $\sim 5$ for the $d$ valence 
quark. It is consistent with our results considering huge uncertainties coming from the radiative corrections and renormalization scaling 
evolution.        

When we consider the time-ordered products for the hard-collinear gluon exchange in the proton, the electromagnetic current should be 
expanded to order $\mO(\lambda^2)$ since all the spectator quarks interact with the active quark. For example we obtain the following 
 operator at $\mO(\lambda^2)$ to give a  leading contribution to the structure function  
\begin{equation} 
J_{\perp}^{(2)\mu} = - \frac{1}{Q^2}\int du_1 du_2  C'_{1} (u_1,u_2) \overline{\Psi}_{\n} \gamma_{\perp}^{\mu} 
\fmsl{B}^{\perp}_{n} \Bigl[\delta(u_2 - \frac{\n\cdot \mP}{\n\cdot P}) \fmsl{B}^{\perp}_{n}\Bigr] 
\Bigl[\delta(u_1 - \frac{\n\cdot \mP}{\n\cdot P}) \Psi_n\Bigr], 
\end{equation}
where $C'_{\perp} (u_1,u_2,\mu)$ is the Wilson coefficient given by $1/(u_1+u_2)$ at tree level and $P^{\mu}$ is the momentum 
of the proton. Since this operator is proportional to $\gamma_{\perp}^{\mu}$, the time-ordered product contributes to $F_1$. 
The anomalous dimension is $C'_1$ is different from $C_1$ in Eq.~(\ref{j01}) and hence we need a new nonperturbative function 
different from the standard PDF $f_{q/p}$, which is induced from the time-ordered products of the leading electromagnetic current
 $J_{\perp}^{(0)\mu}$.

Even though $F_L$ is comparable to $F_1$ in the power counting of $(1-x)$, the precise estimate on the size should include the 
radiative corrections and the evolution of the operators. The dominant contribution to $F_1$ comes from the part involving 
$f_{q/p}$, which is regarded as totally nonperturbative because the factorized expression $f_{q/p} = \mJ\otimes \mS \otimes \Phi$ 
is not justified. When $Q^2$ is large, $\alpha_s (Q^2)$ or $\alpha_s (Q^2(1-x))$ are significantly small. In this case, factorizable parts can be 
considered to be higher order in $\alpha_s$ compared to $f_{q/p}$. If the factorizable contributions are dominant in $F_L$, the size of $F_1$ 
can be larger than $F_L$, which needs to be verified from experiment. For nonleptonic $B$ decays, a similar comparison can be performed
using experimental data. \cite{Bauer:2004tj} For an initial pion, we have seen that $F_L$ is totally factorizable both for hard-collinear and 
for hard gluon exchanges. But for a proton, a more detailed analysis is necessary in order to compare the size of $F_1$ and  $F_L$ 
in the endpoint region. 

\section{Drell-Yan process near the endpoint\label{dyp}}

Near the endpoint in DY process with $1-x_1 \sim 1-x_2 \sim \eta$, the quantity $\tau = Q^2/s$ approaches 1 with the 
power counting $1-\tau \sim \eta$, where $Q^2$ is the invariant mass of the lepton pair and $s$ is the hadronic center-of-mass energy. 
The variables $x_1$ and $x_2$ are defined as
\begin{equation}
x_1 =\frac{Q^2}{2P_1 \cdot q}, \ x_2 =\frac{Q^2}{2P_2\cdot q},
\end{equation}
where $P_1$ and $P_2$ are the momenta of incoming hadrons.
In this limit, the final-state invariant mass becomes 
\begin{equation}
p_X^2 = Q^2 \Bigl(1+\frac{1}{\tau} -\frac{1}{x_1} -\frac{1}{x_2}\Bigr) \rightarrow Q^2(1-x_1)(1-x_2) \sim \Lambda^2,
\end{equation}
requiring that only soft particles be allowed in the final state. 

Since the phase space in this endpoint region is so small, it is not interesting experimentally, but it is a good region to study the
factorization property theoretically. To increase the available phase space, we may think of relaxing the condition such that 
$p_X^2 \sim Q\Lambda$. This region can be reached if only one parton is near the endpoint region, say, $1-x_1\sim 1$ and $1-x_2 \sim \eta$. 
However, since the scattering cross section is a convolution with respect to $x_1$ and $x_2$, it is also possible to have 
$1-x_1, \ 1-x_2 \sim \sqrt{\eta}$ such that $(1-x_1)(1-x_2)\sim \eta$, which corresponds to none of the endpoint region. 
Actually, the region both away from the endpoint region is favored compared to the case with one parton 
near the endpoint region due to the steep decrease of the
PDF near the endpoint. This region might be interesting on its own, but we confine to the above endpoint region here.

The differential scattering cross section is given by 
\begin{equation}
\label{difsca} 
\frac{d\sigma (H_1 H_2 \to l^+l^- X)}{dQ^2} = \sum_f Q_f^2 \frac{2\alpha^2}{3Q^2 s} \frac{1}{4}\sum_{\mathrm{spins}} F_{DY},
\end{equation}
where $F_{DY}$ is the DY structure function, which is given by \cite{Bauer:2002nz}
\begin{equation} 
\label{FDY}
F_{DY} = -\int\frac{d^4 q}{(2\pi)^3} \theta (q^0)  \delta(q^2 - Q^2) \int d^4 z e^{-iq\cdot z} 
\langle H_1 H_2 | J^{\dagger\mu} (z) J_{\mu} (0) | H_1 H_2 \rangle. 
\end{equation}
Here $J^{\mu}$ is an electromagnetic current and the momentum $q$ is given by $q=P_1+P_2-p_X$, where $P_{1,2}$ are the momenta 
of two incoming hadrons $H_{1,2}$.  
In the power counting the product of the volume elements $d^4 z$ and $d^4 q$ yields order 1 irrespective of whether the region is near or away
from the endpoint. Near the endpoint, the label momenta, when integrated over the momentum, yields a Kronecker delta, and the remaining 
$d^4 q$ is of order $\Lambda^4$, while the volume element is of order $\Lambda^{-4}$.  And away from the endpoint, $d^4 q \sim Q^4$, 
and $d^4 z \sim 1/Q^4$. However, there is a delta function
$\delta(q^2-Q^2)$, which is power counted as $\mathcal{D} \sim 1/(Q^2\eta)$ since the argument in the delta function
is given by $q^2 -Q^2 =s(1-\tau)(1-2p_X^0 s^{-1/2}/(1-\tau))$ of order $Q^2 \eta$ in the center-of-mass frame. 

\begin{figure}[b] 
\begin{center}
\includegraphics[height=6cm]{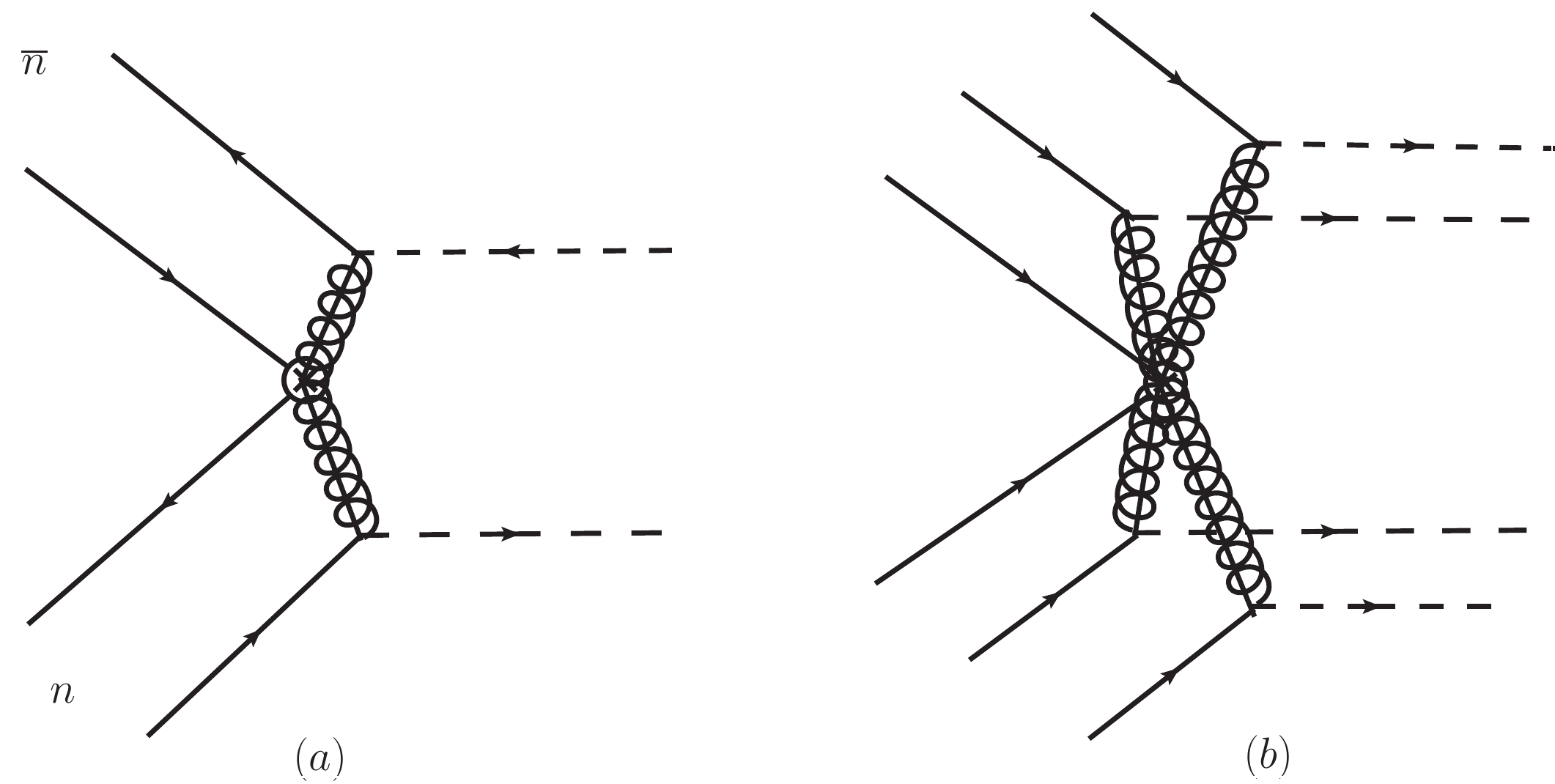}
\caption{\baselineskip 3.0ex 
Examples of the Feynman diagrams with the hard-collinear gluon exchanges for initial (a) pions and (b) proton and antiproton near 
the endpoint, in which the spectator quarks become soft.
\vspace{-.8cm} }
\label{dy}
\end{center}
\end{figure}

We first consider the power counting of the hard-collinear gluon exchange contributions, and some examples of the contributing 
Feynman diagrams are shown in Fig.~\ref{dy}. Since there should be only soft particles in the final state, hard-collinear gluon exchange 
is needed for each final soft quark. Following the same power counting rule as in DIS, the hard-collinear contribution to the structure 
function, namely $F_{DY}^{HC}$ is power counted as
\begin{eqnarray}
\label{hcgdy} 
F_{DY}^{HC} &\sim& \mathcal{D} \cdot \mM \cdot I \cdot F \\
&\sim&  \left\{
\begin{array}{ll} \displaystyle 
\frac{1}{Q\Lambda} \cdot \Bigl(\frac{1}{Q^4\Lambda^2}\Bigr)^2 \cdot (Q\Lambda)^4 \cdot (\Lambda^3)^2  \sim \eta^5 \sim (1-\tau)^5&\mathrm{for}~H_{1,2} = \pi, \\
\displaystyle \frac{1}{Q\Lambda} \cdot \Bigl(\frac{1}{Q^8\Lambda^4}\Bigr)^2 \cdot (Q^3\Lambda^4)^2 \cdot (\Lambda^3)^4  
\sim \eta^{11} \sim (1-\tau)^{11}&\mathrm{for}~H_{1,2} = p,\bar{p}.
\end{array}
\right. \nonumber
\end{eqnarray}
where $F$ is the power-counting on the final soft quark states. 

\begin{figure}[b] 
\begin{center}
\includegraphics[height=5cm]{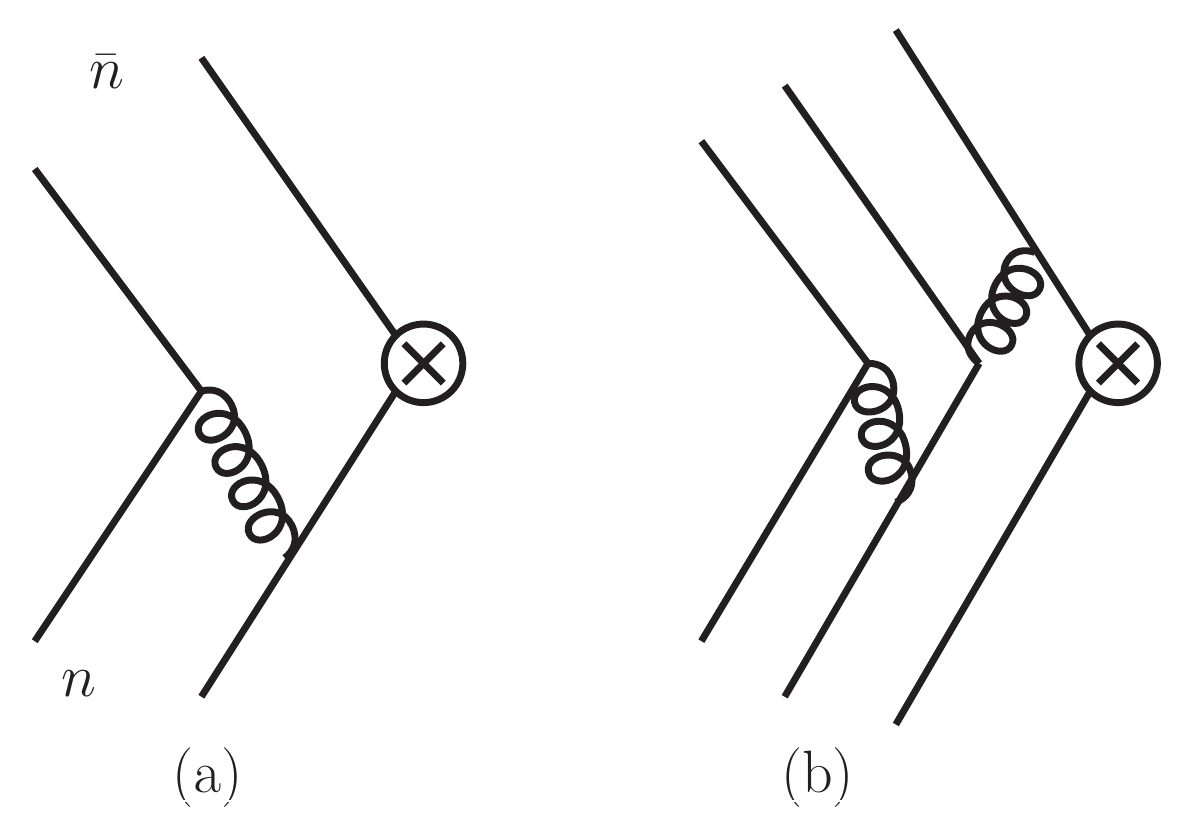}
\caption{\baselineskip 3.0ex 
Feynman diagrams for double parton annihilation in Drell-Yan processes with hard gluon exchange between initial-state 
(a) pions, and (b) proton and antiproton. The diagrams with the gluons attached to other fermions connected to $\otimes$  are omitted.
\vspace{-.5cm}}
\label{dyhard}
\end{center}
\end{figure}

In DY processes, there is no analog of final-state collinear particles in DIS. However, there is another interesting process to be considered
as far as the power counting is concerned. That process is ``double parton annihilation", in which two quark-antiquark pairs in the incoming
hadrons are annihilated by exchanging the momentum of order $Q^2$. This process is shown in Fig.~\ref{dyhard}. The spectator 
quark-antiquark pair with energy fractions of order 1 is annihilated and transfers the whole energy to one of the active quarks. This is similar 
to the case with double parton scattering, but there is the difference in the final states in double parton annihilation. Since the momentum transfer
is of order $Q^2$, the resultant operators become local. Furthermore, they are lower in  powers of $\alpha_s$ compared to 
the corresponding hard-collinear gluon exchanges. That is, these contributions are of order $\alpha_s^2 (Q^2)$ and $\alpha_s^4 (Q^2)$ 
at leading order for pions and (anti)protons respectively.  Using the power counting analysis, the power counting of the structure 
function for initial pions and  $p$, $\overline{p}$ is summarized as
\begin{eqnarray}
\label{hgdy} 
F_{DY}^{H} &\sim& \mathcal{D} \cdot \mM \cdot I  \\
&\sim&  \left\{
\begin{array}{ll} \displaystyle 
\frac{1}{Q\Lambda} \cdot \Bigl(\frac{1}{Q^3}\Bigr)^2 \cdot (Q\Lambda)^4   \sim \eta^3 \sim (1-\tau)^3
&\mathrm{for}~H_{1,2} = \pi, \\
\displaystyle \frac{1}{Q\Lambda} \cdot \Bigl(\frac{1}{Q^6}\Bigr)^2 \cdot (Q^3\Lambda^4)^2  \sim \eta^7 \sim (1-\tau)^7
&\mathrm{for}~H_{1,2} = p,\bar{p}.
\end{array}
\right. \nonumber
\end{eqnarray}
The Feynman diagrams in Fig.~\ref{dyhard} can be dressed with soft gluons for final-state soft particles, but careful analysis of power counting
shows that emission of soft gluons does not alter the result of the power counting without soft gluons.
One thing to note in Fig.~\ref{dyhard} (a) is that the Feynman diagram, when rotated, is exactly the same as the one for the pion form factor. 
It is interesting that the pion form factor and the double parton annihilation in DY processes are related.

The complication in DY processes near the endpoint lies in the fact that there exists no limiting process from the conventional approach, and
the double parton annihilation is less suppressed both in powers of $\alpha_s$ and $1-\tau$.
Among the contributions from hard-collinear gluon exchange, there can be nonfactorizable contributions when we take the LCDA for 
the initial state. If these nonfactorizable contributions are dominant, we can arguably regard $F_{DY} \sim (1-\tau)^5$ or $(1-\tau)^{11}$ 
from hard-collinear exchange
without additional suppression by multiple powers of $\alpha_s(Q\Lambda)$, as we considered on the estimate of the sizes of $F_1$ and $F_L$
in DIS. In that case, these contributions from hard-collinear gluon exchange 
can be numerically  comparable to the hard gluon contributions resulting in double parton annihilation,  treating $\alpha_s (Q^2) \sim 1-\tau$.  
On the other hand, if the double parton annihilation is the major contribution near the endpoint region, its effect may be noticeable as 
we get away from the endpoint
region. But note that the conventional leading contribution of order 1 becomes dominant away from the endpoint region, and all the 
contributions considered 
above become subleading and are negligible. In some region between the standard region and the endpoint region, the effect of the double parton
annihilation may be noticeable. However,  for  precise estimate and comparison, a more detailed analysis is necessary.    

The conventional approaches neglecting the spectator contribution have proposed the following factorization formula~\cite{Sterman:1986aj,Catani:1989ne,Idilbi:2005ky,Becher:2007ty}
\begin{equation}
\label{cfdy} 
F_{DY} = H_{DY} (Q^2) \int^1_{\tau} \frac{dz}{z} S_{DY} (1-z) f_{DY} \Bigl(\frac{\tau}{z}\Bigr),
\end{equation}
where $H_{DY}$ is the hard function of order 1,  $S_{DY}$ is the soft function consisting of the products of the soft Wilson lines, 
and $f_{DY}$ is  the convolution of the parton distributions, which is given by 
\begin{equation}
\label{dyparton}
f_{DY} \Bigl(\frac{\tau}{z} \Bigr) = \int^1_{\tau/z} \frac{dy}{y} f_{q/H_1} (y) f_{\bar{q}/H_2} \Bigl(\frac{\tau}{zy} \Bigr).
\end{equation} 
Since $\int dz S_{DY} (1-z)$ in Eq.~(\ref{cfdy}) is of order 1. The power counting of the structure function in the conventional 
approach can be performed through $f_{DY}$. Since $f_{q/H}$ scales as $(1-x)^2$ for the pion and $(1-x)^5$ for the proton in DIS
according to our analysis, $F_{DY}$ can be power-counted as $(1-\tau)^5$ or $(1-\tau)^{11}$ treating the range of the integration 
in Eq.~(\ref{dyparton}) to be of order $\eta$. Therefore the estimate of the size in the conventional approach seems to favor the power 
counting of the hard-collinear contribution in Eq.~(\ref{hcgdy}).  However it is not clear whether we can justify the parameterization 
of the contributions from hard or hard-collinear gluon exchanges as the convolution of the PDFs. 

\section{Conclusion\label{conc}}
High-energy scattering processes near the endpoint region are hard to analyze in experiment, but they offer an intriguing opportunity 
to disentangle the structure of factorization properties in QCD. In this paper, a power counting analysis is performed for the structure 
functions in DIS and in DY processes near the endpoint region to claim that there are new contributions from hard-collinear gluon 
exchanges to be included since they are comparable to the currently available leading contributions. 

An important feature in this analysis is to apply kinematic constraints of the endpoint region to classify the possible types of
final-state particles, while the initial partons and hadrons are required to be on the mass shell $p^2 \sim \Lambda^2$.  The resonance region 
is defined as the final states with $p^2 \sim \Lambda^2$, and the endpoint region is defined as those with $p^2 \sim Q\Lambda$. 
According to this classification, DIS can have both the resonance region and the endpoint region, but DY processes have actually only 
the resonance region.

The explicit factorization proof for hard-collinear gluon exchanges in DIS is interesting in itself, but it is also  illuminating to compare 
this process with nonleptonic $B$ decays into two light mesons. In the factorization proof for nonleptonic $B$ decays \cite{Chay:2003ju}, 
we have considered the contribution of the four-quark operators along with the spectator interactions since they are of the same order.  
In the spectator interaction, a hard-collinear gluon is exchanged between the four-quark operator and a spectator quark in a $B$ meson, 
and the final-state particles become collinear to form mesons. The hard-collinear gluon exchange considered here in DIS  is exactly the reverse 
process of this spectator interaction, in which the final-state collinear particles are the incoming partons, and the initial soft quark is 
the final soft particle, and the heavy $b$ quark is replaced by the $\overline{n}$-collinear final-state jet. The factorization property of 
various spectator interactions is similar in both cases,  noting the difference between a heavy quark and an $\overline{n}$-collinear particle.
This, along with the comparison between the double parton annihilation in DY processes and the pion form factor, shows an interesting 
relationship among different processes.

In DIS, the spectator interaction has the same power counting as the process with final $\overline{n}$-collinear particles, hence it should be
included to be consistent. However, in DY processes, the spectator interaction exists, but it is suppressed compared to the double parton 
annihilation. This result is surprising, but here we have considered only the power counting of various contributions, and we have not 
tried to give numerical analysis of those since it belongs to a future work. The power counting analysis indicates the degree of 
suppression in powers of $1-x$ or $1-\tau$, but the actual contributions also involve other parameters such as some powers of 
$\alpha_s$ at different scales $Q^2$, $Q\Lambda$. Therefore a study on the precise estimate of various contributions is necessary 
to compare with experiment.

\begin{acknowledgments}
J. C. is supported  by Mid-career Researcher Program through NRF grant funded by the MEST (2009-0086383). 
Both authors are supported  by Basic Science Research Program through the NRF of Korea funded by the MEST (2009-0072611).  
\end{acknowledgments}

\end{document}